\documentclass[conference]{IEEEtran}
\IEEEoverridecommandlockouts
\usepackage{cite}
\usepackage{amsmath,amssymb,amsfonts}
\usepackage{algorithmic}
\usepackage{graphicx}
\usepackage{textcomp}
\usepackage{xcolor}
\def\BibTeX{{\rm B\kern-.05em{\sc i\kern-.025em b}\kern-.08em
    T\kern-.1667em\lower.7ex\hbox{E}\kern-.125emX}}
\begin{document}

\title{Multi-Beam Automotive SAR Imaging \\in Urban Scenarios}

\author{

\IEEEauthorblockN{Marco Rizzi}
\IEEEauthorblockA{\textit{DEIB} \\
\textit{Politecnico di Milano}\\
Milano, Italy \\
marco.rizzi@polimi.it}
\and

\IEEEauthorblockN{Marco Manzoni}
\IEEEauthorblockA{\textit{DEIB} \\
\textit{Politecnico di Milano}\\
Milano, Italy \\
marco.manzoni@polimi.it}
\and

\IEEEauthorblockN{Stefano Tebaldini}
\IEEEauthorblockA{\textit{DEIB} \\
\textit{Politecnico di Milano}\\
Milano, Italy \\
stefano.tebaldini@polimi.it}
\and

\IEEEauthorblockN{Andrea Virgilio Monti-Guarnieri}
\IEEEauthorblockA{\textit{DEIB} \\
\textit{Politecnico di Milano}\\
Milano, Italy \\
andrea.montiguarnieri@polimi.it}
\and

\IEEEauthorblockN{Claudio Maria Prati}
\IEEEauthorblockA{\textit{DEIB} \\
\textit{Politecnico di Milano}\\
Milano, Italy \\
claudio.prati@polimi.it}
\and

\IEEEauthorblockN{Dario Tagliaferri}
\IEEEauthorblockA{\textit{DEIB} \\
\textit{Politecnico di Milano}\\
Milano, Italy \\
dario.tagliaferri@polimi.it}
\and

\IEEEauthorblockN{Monica Nicoli}
\IEEEauthorblockA{\textit{DIG} \\
\textit{Politecnico di Milano}\\
Milano, Italy \\
monica.nicoli@polimi.it}
\and

\IEEEauthorblockN{Ivan Russo}
\IEEEauthorblockA{\textit{Huawei Milan Research Center} \\
Huawei Technologies Italia S.r.l.\\
Segrate, Italy \\
ivan.russo@huawei.com}
\and

\IEEEauthorblockN{Christian Mazzucco}
\IEEEauthorblockA{\textit{Huawei Milan Research Center} \\
Huawei Technologies Italia S.r.l.\\
Segrate, Italy \\
christian.mazzucco@huawei.com}
\and

\IEEEauthorblockN{Simón Tejero Alfageme}
\IEEEauthorblockA{\textit{Huawei German Research Center}\\
Huawei Technologies Duesseldorf GmbH \\
Munich, Germany \\
simon.tejero.alfageme@huawei.com}
\and

\IEEEauthorblockN{Umberto Spagnolini}
\IEEEauthorblockA{\textit{DEIB} \\
\textit{Politecnico di Milano}\\
Milano, Italy \\
umberto.spagnolini@polimi.it}

}

\maketitle

\begin{abstract}
Automotive synthetic aperture radar (SAR) systems are rapidly emerging as a candidate technological solution to enable a high-resolution environment mapping for autonomous driving. Compared to lidars and cameras, automotive-legacy radars can work in any weather condition and without an external source of illumination, but are limited in either range or angular resolution. SARs offer a relevant increase in angular resolution, provided that the ego-motion of the radar platform is known along the synthetic aperture. In this paper, we present the results of an experimental campaign aimed at assessing the potential of a multi-beam SAR imaging in an urban scenario, composed of various targets (buildings, cars, pedestrian, etc.), employing a 77 GHz multiple-input multiple-output (MIMO) radar platform based on a mass-market available automotive-grade technology. The results highlight a centimeter-level accuracy of the SAR images in realistic driving conditions, showing the possibility to use a multi-angle focusing approach to detect and discriminate between different targets based on their angular scattering response. 
\end{abstract}


\section{Introduction}

The rush to fully-autonomous vehicles requires the usage of a huge heterogeneous set of different sensors, such as cameras, lidars, radars, acoustic, etc. The range of functionalities is wide, from basic emergency braking to advanced environmental perception \cite{Marti2019ADAS_sensors}. Cameras and lidars are, respectively, passive and active optical sensors able to create high-resolution images and/or point clouds of the surrounding, that suitably integrated provide the vehicles with the capability of detect and classify objects in the environment. Among the disadvantages, the most important is the sensitivity to adverse weather conditions (rain, fog, etc.) and the requirement of an external source of illumination (cameras), challenging their usage for safety critical applications. In this regard, automotive-legacy multiple-input multiple-output (MIMO) radars working in the 76-81 GHz band (W-band)~\cite{ETSI_TR103593} are widely employed to obtain measures of radial distance, velocity and angular position of remote targets at short, medium or long range (up to 250 meters), but are characterized by a poor trade-off between hardware cost, angular resolution (typically $> 1$ deg), bandwidth (typically around 600 MHz for medium-long range radars) and field of view (FoV) \cite{Hasch2012,Brisken2018}. Although radar-based simultaneous localization and mapping works are present in literature \cite{Holder2019}, synthetic aperture radar (SAR) techniques were increasingly used since the last 10 years to augment the accuracy of environmental perception \cite{Zwick2009_SARforparking,Iqbal2015_SARforparking,Feger2017_experimentalSAR77GHz,Stanko2016_Miranda,Tagliaferri2021_SARnavigation}. The aim is to perform a radio imaging of both static and non-static targets (pedestrians, vehicles, buildings, etc.) of the driving environment, i.e., generating a two- or three-dimensional spatial map of a given scene where the targets appear at pixel locations that correspond to their physical positions in space, adding detailed information about the shape of the illuminated objects. 
Early works in \cite{Zwick2009_SARforparking,Zwick2011_motioncompSARgyroacc} were targeted to a simple parking lot detection, with either simulations or  in-laboratory/preliminary outdoor experiments using a 24 GHz frequency-modulated continuous waveform (FMCW) radar to assess the feasibility of the application. Relevance was given to the issue of knowing the position of the radar platform along its motion and preliminary countermeasures. Similarly, in \cite{Iqbal2015_SARforparking} the authors performed a measurement campaign with a side-looking SAR mounted on a linear rail unit, with the aim of imaging two parked cars.
In \cite{Feger2017_experimentalSAR77GHz}, an experimental side-looking SAR system, working at 77 GHz, is mounted on the vehicle’s rooftop, observing the front-right quadrant of the car., performing the imaging of the surrounding at a resolution of 15 cm. Further investigations are in \cite{Laribi2018} and \cite{Wang2019_parkingSAR}, respectively. An example of millimeter-accurate SAR imaging is provided in \cite{Stanko2016_Miranda}, where an expensive setup comprising a 300 GHz radar with 40 GHz bandwidth is mounted on a van, slowly travelling along a perfectly linear path, producing high-resolution images of street markings, objects and people. More recent works were aimed at addressing the real-time challenge for automotive SAR systems \cite{Kan2020_realtimeSAR,Gisder2019_automotiveSAR_wheelspeed_CTRV, Gisder2018_automotiveSAR_wheelspeed,Iqbal2021}, suggesting the usage of radar-only motion compensation and dedicated graphics processing units to enable massive parallel computing of the backprojection integral.   

In this paper, we stem from the current state of the art and from our previous work \cite{Tagliaferri2021_SARnavigation} to investigate the performance of a multi-beam SAR focusing approach in urban scenarios. A preliminary work in this direction was carried out in \cite{Harrer2018_multichannel}, where the authors tested a stripmap, squinted and spotlight SAR mode for parking lot detection only. Here, we are interested in the high-resolution detection and discrimination of \textit{different} targets of a typical urban scenario, with particular interest in pedestrians.  Experimental results, obtained from a dedicated data acquisition campaign with a fully-equipped car, reveal the possibility of imaging different targets (buildings, fences, pedestrians, parked cars, etc.) at a centimeter-level accuracy, based on their peculiar angular response at 77 GHz. In particular, this last result could increase the robustness and the reliability of camera- and lidar-based object detection and classification in the context of autonomous driving. 

The remainder of the paper is organized as follows: Section \ref{sect:campaign} describes the experimental campaign in the selected urban scenario, Section \ref{sect:processing} details the navigation and SAR data processing on the collected data, Section \ref{sect:results} outlines and discusses the results while Section \ref{sect:conclusion} draws the conclusions.

\section{Acquisition Campaign}\label{sect:campaign}
\begin{figure}[!t]
    \centering
    \includegraphics[width=\columnwidth]{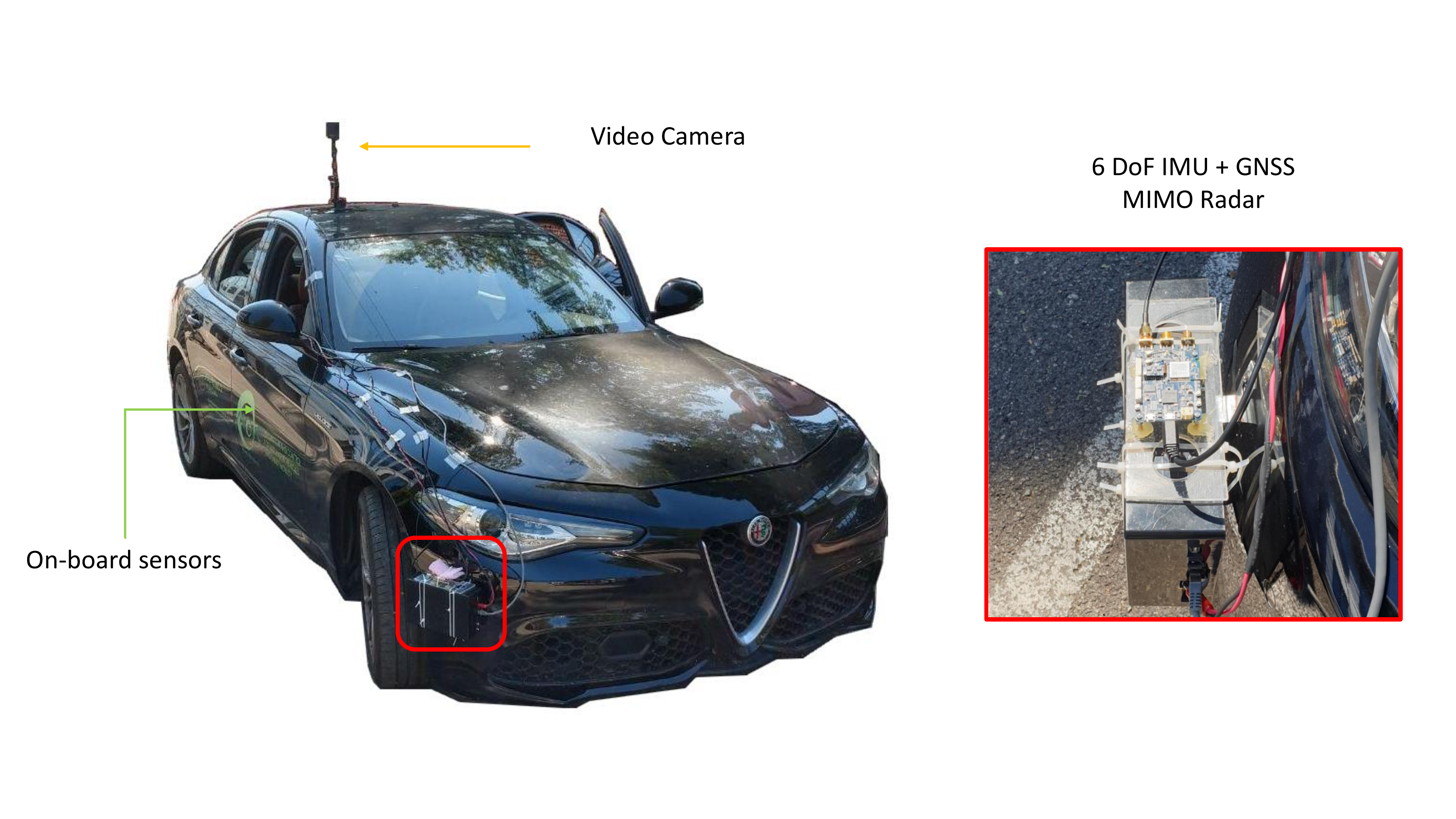}
    \caption{Alfa Romeo Giulia sensor equipment. In the red box: IMU+GNSS dedicated module on top of Scanbrick\textsuperscript{\textregistered} W43 in forward squint-looking configuration.}
    \label{fig:Alfa_car}
\end{figure}

A dedicated experimental campaign was carried out to gather the radar and navigation data used to demonstrate the potential of multi-beam urban SAR imaging. We employed a fully equipped Alfa Romeo Giulia ``Veloce'', depicted in Fig. \ref{fig:Alfa_car}, equipped with an external ScanBrick\textsuperscript{\textregistered} W43 radar, see Table \ref{SBparameters}, (proprietary platform by Aresys) rigidly mounted on the front bumper of the vehicle, about $0.5$ m above ground, and pointed at $60$ degrees with respect to the driving direction, in a frontal-side looking configuration. The ScanBrick\textsuperscript{\textregistered} is a short range FMCW $3\times 4$ multiple-input multiple-output (MIMO) radar operating in the 77-81 GHz band (3 GHz of bandwidth) with a pulse repetition frequency (PRF) of $990$ Hz, and it is based on a mass-market available commercial automotive technology. 

\begin{table}[h]
\caption{ScanBrick\textsuperscript{\textregistered} W43}
\begin{center}
\begin{tabular}{|c|c|}
\hline
\textbf{Parameter}&\textbf{Value} \\
\hline
Waveform & FMCW \\
\hline
Number of Channels & $8$ \\
\hline
Carrier Frequency& $77$ GHz \\
\hline
Bandwidth & $3$ GHz \\
\hline
Pulse Duration & $155\ \mu$s \\
\hline
PRF & $990$ Hz \\
\hline
Max. Range & $26$ m \\
\hline
Range resolution & $5$ cm \\
\hline
Angular resolution & $14.3$ deg \\
\hline

\end{tabular}
\label{SBparameters}
\end{center}
\end{table}

The Alfa Romeo Giulia car was also purposely equipped with on-board and external navigation sensors to provide the ground truth radar platform motion synthesizing the aperture. The car on-board sensor equipment comprises: \textit{(i)} two co-located 3 degrees of freedom (DoF) inertial measurement units (IMUs), measuring both lateral and longitudinal acceleration, and heading rate; \textit{(ii)} a 6 DoF IMU plus a global navigation satellite system (GNSS) module from Suchy Data Systems GmbH \cite{SuchyxProGPSnano} in the vehicle center of gravity (CoG), measuring the 3D acceleration and the 3D angular velocity; \textit{(iii)} a 3 DoF IMU in the rear part of the car; \textit{(iv)} four wheel velocity sensors; \textit{(v)} a steering angle sensor at the frontal wheels. Furthermore, a SAR-dedicated 6 DoF IMU+GNSS integrated sensor, from Inertial Sense \cite{InertialSense2020}, is rigidly mounted on top of the Scanbrick\textsuperscript{\textregistered} radar (Fig. \ref{fig:Alfa_car}). Finally, a video camera has been placed on the vehicle rooftop to cross-check the acquired data.

\addtolength{\textfloatsep}{-4mm}

\begin{figure*}[!t]
    \centering
    \includegraphics[width=1.6\columnwidth]{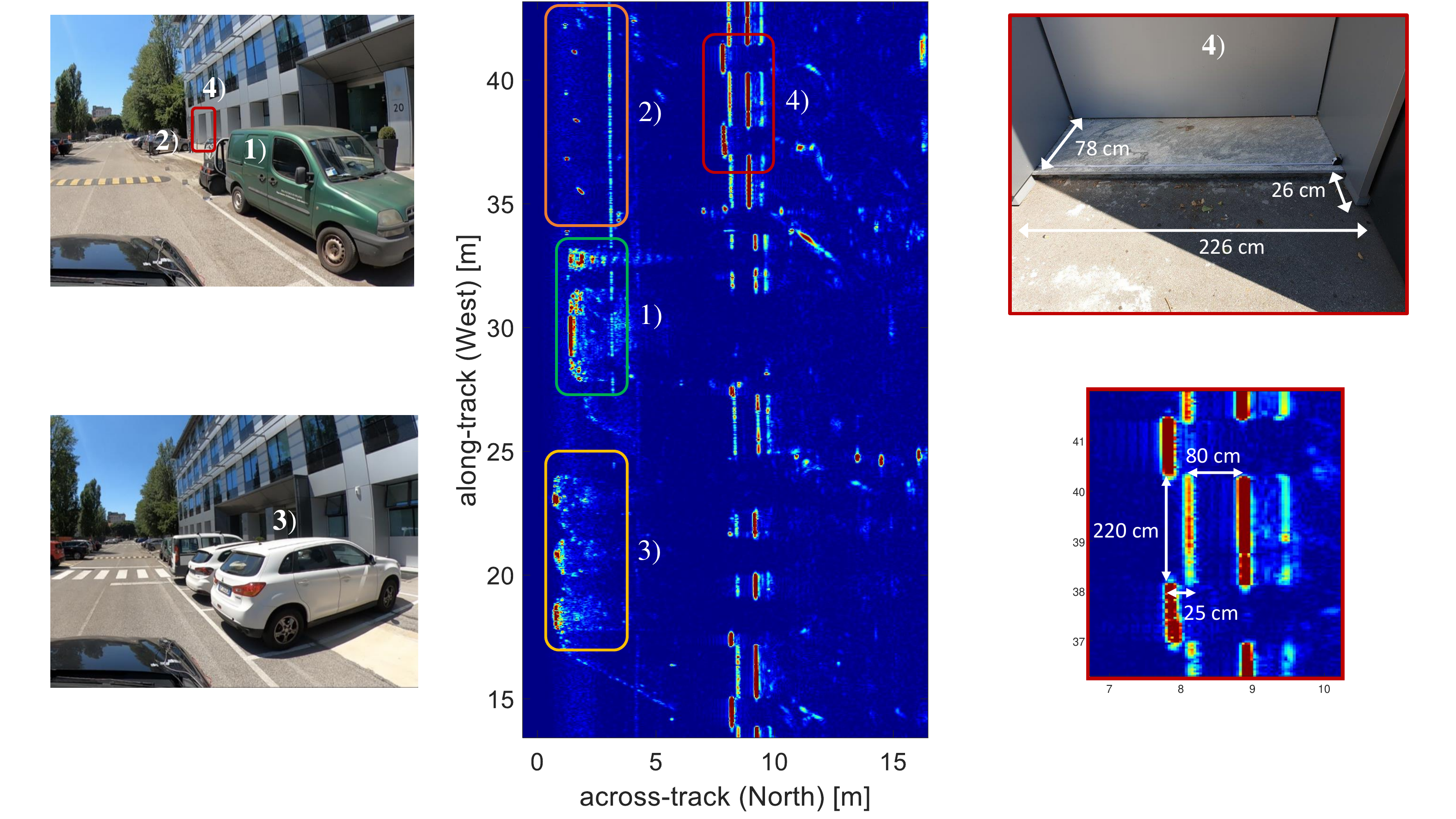}
    \caption{Test 1: Single beam SAR image target identification (image not in scale).}
    \label{fig:Test_5}
\end{figure*}
\begin{figure*}[!t]
    \centering
    \includegraphics[width=1.6\columnwidth]{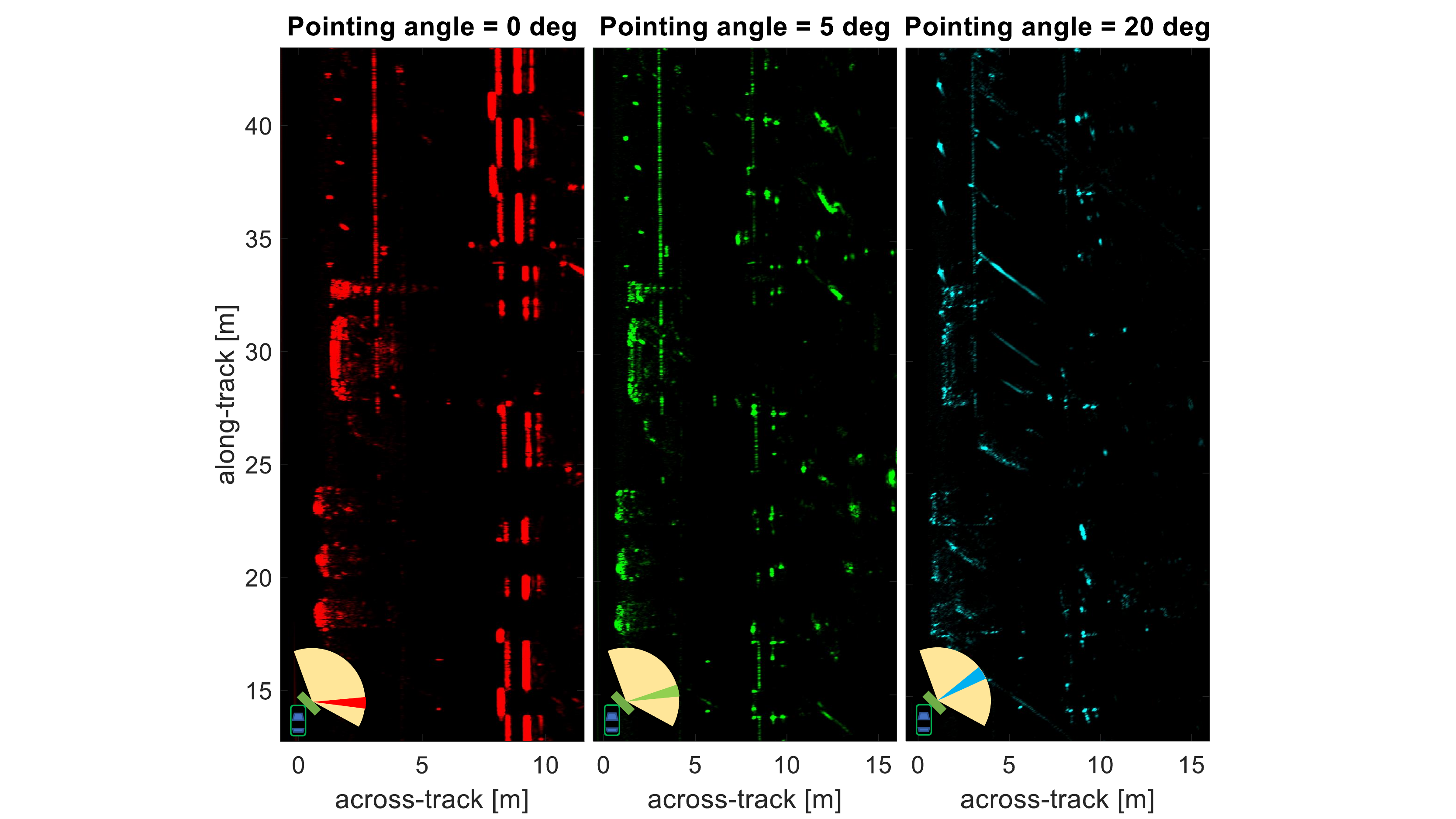}
    \caption{Test 1: set of single-beam SAR images RGB coded: 0 deg in red, 5 deg in green, 20 deg in light blue (images not in scale).}
    \label{fig:Test_5_angle}
\end{figure*}

\section{Data Processing}\label{sect:processing}

In the following, we briefly describe the navigation and radar data processing applied to experimental data.

\subsection{Navigation Data Processing}
Navigation data from the on-board sensors are fused with an unscented kalman filter (UKF) \cite{Wan2000UKF} to track the 2D radar position, the velocity and the heading over time. We choose the constant turn rate and acceleration model to describe the state evolution and the driving process over time, that best represents the dynamics experienced in the experimental campaign, characterized by slowly varying speeds and heading rates. The measurements of on-board sensors (Section \ref{sect:campaign}) compose the set of observations for the UKF. Details about the navigation processing can be found in our previous work~\cite{Tagliaferri2021_SARnavigation}.

\subsection{Radar Data Processing}
The radar data processing chain follows the standard procedure of FMCW radars. The range-compressed (RC) data matrix is obtained from raw data through a fast-Fourier transform (FFT) in the fast-time domain. RC data can be modelled as:
\begin{equation}\label{eq:RangeCompressed_data}
\begin{split}
d_{RC}(t,\tau) = \sum_{\ell} \,s_{RC}(t;T^{(\ell)}_D(\tau)),
\end{split}
\end{equation}
function of the fast-time $t$ and slow-time $\tau$, where the reference RC signal is
\begin{equation}\label{eq:RangeCompressed_signal}
    \begin{split}
        s_{RC}(t;T^{(\ell)}_D(\tau))= A^{(\ell)}  T_{p}\, \mathrm{sinc}&\left[B(t-T^{(\ell)}_D(\tau))\right] \times \\ & \times \exp (j2\pi f_{0}T^{(\ell)}_D(\tau))
    \end{split}
\end{equation}
\begin{figure*}[!t]
    \centering
    \includegraphics[width=178mm]{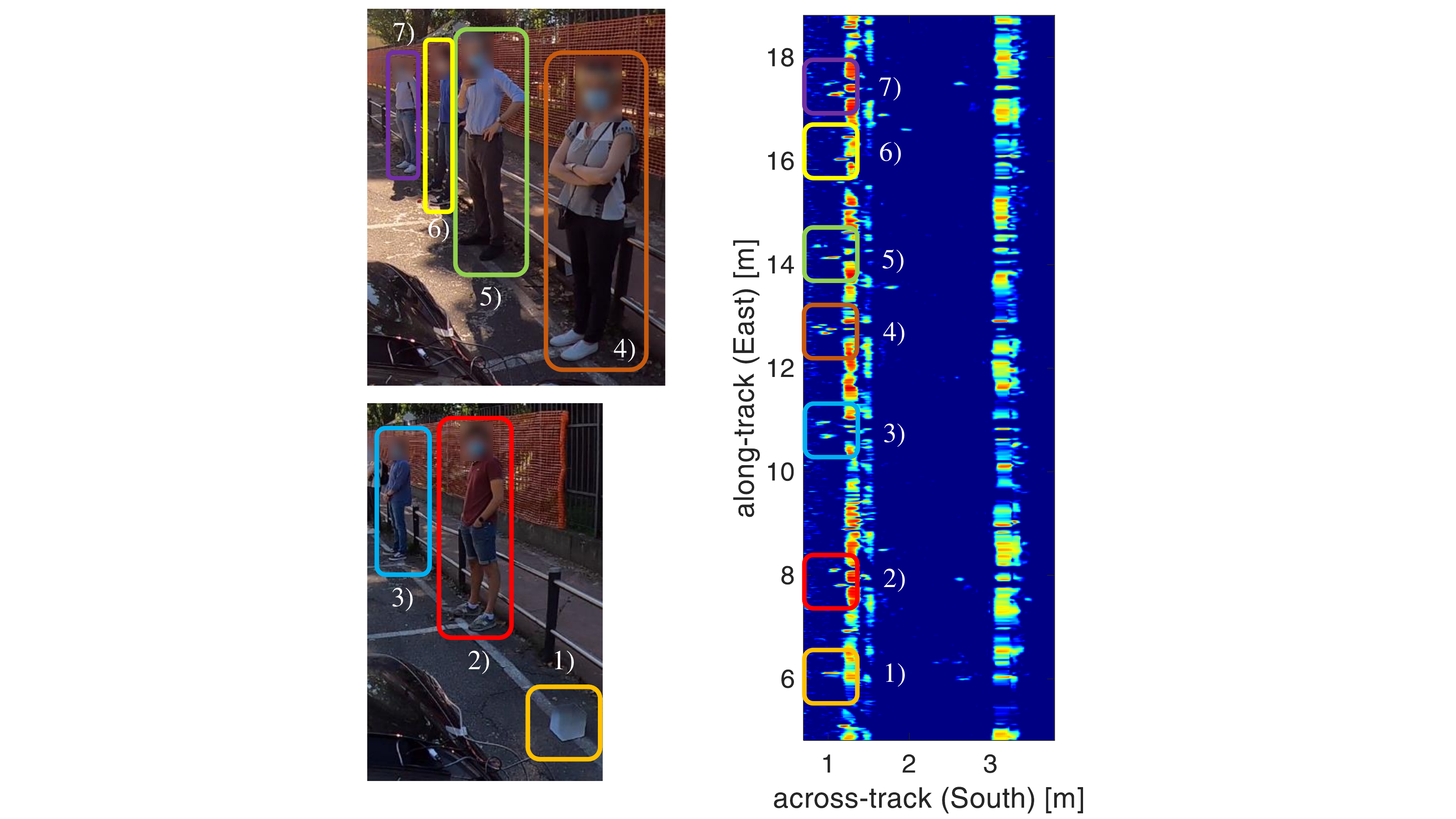}
    \caption{Test 2: Single beam SAR image target identification (image not in scale). 1) corner, 2) through  7) pedestrian targets.}
    \label{fig:Test_3}
\end{figure*}

is the $\ell$-th target RC signal with complex amplitude $A$, for carrier frequency $f_{0}$, bandwidth $B$ and pulse duration $T_p$, characterized by a time-varying two-way delay $T^{(\ell)}_D(\tau)$ and $\mathrm{sinc}\left[x\right] = \sin(x)/x$~\cite{Zaugg2015_FMCWSAR}.
The second phase forms the SAR image.
The main goal of obtaining geometrically accurate maps of the environment motivates the choice of a time-domain back-projection (TDBP) based approach. Although being computationally expensive, the TDBP algorithm yields exact reconstruction of the surroundings for an arbitrary acquisition trajectory \cite{Zaugg2015_FMCWSAR}. The highly non-linear tracks of vehicles in urban scenarios, in terms of direction and velocity, justify the choice of TDBP focusing algorithm. 
To diminish the computational complexity of standard TDBP and reduce the effect of residual motion errors, we take advantage of multi-angle SAR processing. The wide available FoV is limited by enforcing a spatial filter $F(\alpha(\tau;x,y);\alpha_P)$ on the angles of arrival $\alpha$ around the pointing direction of interest $\alpha_P$. The TDBP integral is therefore:
\begin{figure*}[!t]
    \centering
    \includegraphics[width=178mm]{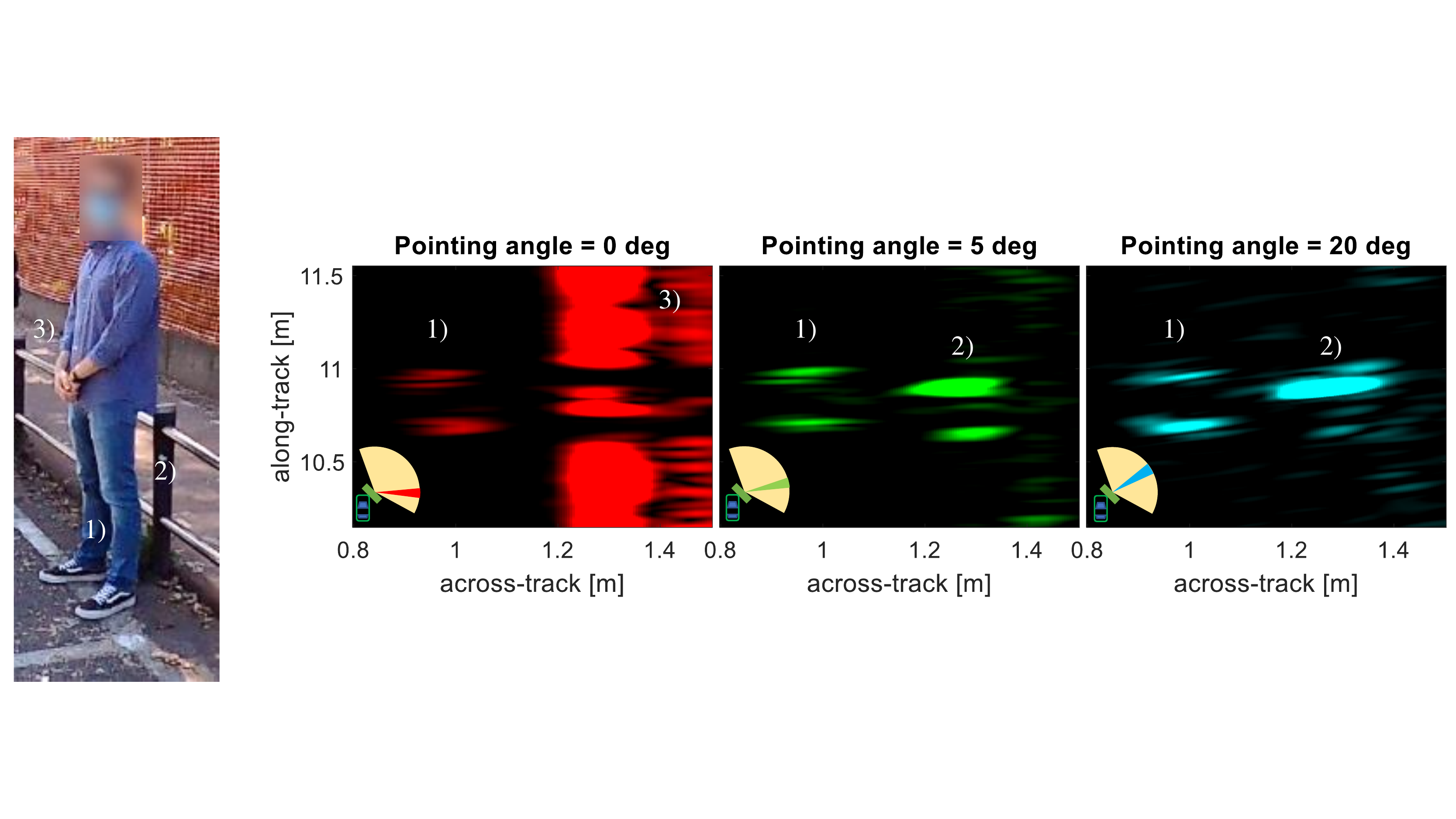}
    \caption{Test 2: Multi-beam SAR images of single person RGB coded: 0 deg in red, 5 deg in green, 20 deg in light blue (images not in scale).}
    \label{fig:Test_3_MR}
\end{figure*}
\begin{equation}\label{eq:stripmapSAR}
\begin{split}
    I(x,y) = \iint & F(\alpha (\tau;x,y);\alpha_P) \,\times \\
    & \times\, d_{RC}(t,\tau) \, s^*_{RC}(t; T_D(\tau;x,y)) \, \, \mathrm{d}t \,\mathrm{d}\tau.
\end{split}
\end{equation}

where $I(x,y)$ is the SAR image and 
\begin{equation}\label{eq:2-wayDelay}
    \begin{split}
    T_{D}(\tau;x,y)=\frac{2}{c}\sqrt{(x-p_{x}(\tau))^{2}+(y-p_{y}(\tau))^{2}}
    \end{split}
\end{equation}
is the two-way propagation delay between any given point of the radar trajectory $\left(p_{x}(\tau),\,p_{y}(\tau)\right)$ (estimated by the UKF) and the image coordinates $(x,y)$ ($c=3\times 10^{8}$ m/s, superscript "$^*$" denotes the complex conjugate).

The range resolution, ruled by the system bandwidth $B=3$ GHz, is $5$ cm and the spatial filter $F(\alpha(\tau;x,y);\alpha_P)$ is here designed such that to achieve the same resolution in cross-range:
\begin{equation}\label{eq:SatialFilter}
    \begin{split}
    F(\alpha(\tau;x,y);\alpha_P)=\exp\bigg\{{-4\Big(\frac{\alpha(\tau;x,y) - \alpha_P}{\Delta_\alpha}\Big)^2}\bigg\}
    \end{split}
\end{equation}
with $\alpha(\tau;x,y)$ being the angles corresponding for each pixel of the grid for the given slow-time index and  $\Delta_\alpha$ is the desired angular resolution of the beam.

\section{Experimental Results}\label{sect:results}
The data were collected on the closed road in front of the Dipartimento di Elettronica, Informazione e Bioingegneria (DEIB) of Politecnico di Milano. The results here presented are selected from two different trajectories of $\approx 50$ m (Test 1) and  $\approx 22$ m (Test 2) length, with a lateral deviation from a perfectly straight motion of around $\pm 1.5$ m. In the first test, the maximum vehicle's speed was constrained to $20$ km/h by the radar PRF and number of channels to obtain unambiguous imaging. The second track was travelled at a maximum velocity of $5$ km/h to guarantee a longer illumination time for the imaging of pedestrian targets. The observed scenes, which are shown in snapshots from the installed video camera, comprise typical urban targets, such as parked cars, sidewalks, fences, buildings, pedestrians as well as reflective corners placed for reference along the road.

\subsection{Test 1}
The focused MIMO SAR image presented in Fig. \ref{fig:Test_5} is obtained by processing a single-beam, specifically the beam perpendicular to the main direction of motion. The side of the mini-van as well as the backside of the electric car, highlighted by detail 1), are clearly distinguishable. Furthermore, the rear of the three parked cars in 2), the series of corners and the sidewalk in 3) are correctly focused. Remarkably, detail 4) highlights the cm-level accuracy achieved in the imaging of the peculiar facade of the DEIB thanks to the ad-hoc ego-motion estimation of the car along the synthetic aperture.
Fig. \ref{fig:Test_5_angle} represents the multi-beam image of the same test. In particular, the three processed beams corresponding to $0$ deg, $5$ deg and $20$ pointing angles are encoded in an Red-Green-Blue (RGB) fashion, respectively, to better distinguish the image content, as sketched in the bottom-left corner of each image of Fig. \ref{fig:Test_5_angle}. Most of the features in the environment, such as cars and corners, are present in all the focused images, while, interestingly, the department wall disappears when focusing beams even slightly off from the $0$ deg pointing angle. Furthermore, different parts of the car are responsible for stronger returns at different angles. At $0$ deg, the main contribution comes from the mini-van door, while for increasing angles, the backscatter from wheel arches region becomes more relevant.

\subsection{Test 2}

The results of Test 2 concerning the usage of a single beam are shown in Fig. \ref{fig:Test_3}. The most important observation is that pedestrians can be effectively detected and recognized with SAR imaging. Specifically, the main scattering belongs to the legs of a person as highlighted in 2) through 7). Furthermore, the corner in 1) and the two metal fences at the side of the sidewalk are correctly reconstructed.
The potential of multi-beam focusing on Test 2 is clear from Fig. \ref{fig:Test_3_MR}. Structures characterized by anisotropic scattering features, such as building and fences, Fig. \ref{fig:Test_3_MR} 3), that are parallel to the road, disappear when the processing beam is pointed differently from the orthogonal one with respect to the direction of motion, while revealing targets with more isotropic scattering properties, that in our case are pedestrian legs 1) and poles 2).

\section{Conclusions}\label{sect:conclusion}

This paper presents experimental results on the potential of automotive SAR imaging for cm-level accuracy mapping of urban scenarios using general-purpose hardware. The employed multi-beam SAR processing turns out to be beneficial for revealing isotropic or nearly isotropic targets (poles, pedestrians), that can be shadowed by strong directional scatterers parallel to the road (walls, fences, etc.). The results suggest the usage of multi-beam SAR as a complementing technology for robust target recognition in all-weather automated driving applications.
Future works include the implementation of more efficient focusing algorithms in the perspective of real-time imaging, along with fast autofocus routines for correction of residual motion errors.


\section*{Acknowledgment}

The research has been carried out in the framework of the Huawei-Politecnico di Milano Joint Research Lab on automotive SAR. The Authors want to acknowledge Dr. Paolo Falcone from Aresys for the cooperation and support in the data acquisition campaign.

\bibliographystyle{IEEEtran}
\bibliography{Bibliography}

\end{document}